\documentclass[aps, prl,twocolumn, showpacs, superscriptaddress,preprintnumbers, amsmath, epsfig, floatfix,normalem]{revtex4-1}

\usepackage{graphicx}
\usepackage{dcolumn}
\usepackage{bm}
\usepackage{ulem}

\usepackage[utf8]{inputenc}
\usepackage[english]{babel}
\usepackage{keyval}
\graphicspath{{figs/}{figs:}{\figs}}
\setkeys{Gin}{width=0.90\columnwidth}

\usepackage[usenames,dvipsnames]{color}
\usepackage[]{ulem}

\newcommand{\comment}[1]{\textcolor{red}{#1}}
\renewcommand{\comment}[1]{\relax}

\newcommand{\todelete}[1]{\textcolor{green}{\sout{#1}}}
\renewcommand{\todelete}[1]{\relax}

\begin{document}

\title{Near room-temperature colossal magnetodielectricity and multiglass properties in partially-disordered La$_2$NiMnO$_6$}
\date{\today}

\author{D. Choudhury}
\affiliation{Solid State and Structural Chemistry Unit, Indian Institute of Science, Bangalore-560012, India}
\affiliation{Department of Physics, Indian Institute of Science, Bangalore-560012, India}

\author{P. Mandal}
\affiliation{Chemistry and Physics of Materials Unit, Jawaharlal Nehru Centre for Advanced Scientific Research, Bangalore - 560064, India}

\author{R. Mathieu}
\affiliation{Department of Engineering Sciences, Uppsala University, SE-75121, Uppsala, Sweden}

\author{A. Hazarika}
\affiliation{Solid State and Structural Chemistry Unit, Indian Institute of Science, Bangalore-560012, India}

\author{S. Rajan}
\affiliation{Solid State and Structural Chemistry Unit, Indian Institute of Science, Bangalore-560012, India}

\author{A. Sundaresan}
\affiliation{Chemistry and Physics of Materials Unit, Jawaharlal Nehru Centre for Advanced Scientific Research, Bangalore - 560064, India}

\author{U. V. Waghmare}
\affiliation{Theoretical Sciences
 Unit, Jawaharlal Nehru Centre for Advanced Scientific Research, Bangalore - 560064, India}

\author{R. Knut}
\affiliation{Department of Physics and Astronomy, Uppsala University, Box-516, SE-75120, Uppsala, Sweden}

\author{O. Karis}
\affiliation{Department of Physics and Astronomy, Uppsala University, Box-516, SE-75120, Uppsala, Sweden}

\author{P. Nordblad}
\affiliation{Department of Engineering Sciences, Uppsala University, SE-75121, Uppsala, Sweden}

\author{D. D. Sarma$^{*,}$}
\affiliation{Solid State and Structural Chemistry Unit, Indian Institute of Science, Bangalore-560012, India}
\affiliation{Department of Physics and Astronomy, Uppsala University, Box-516, SE-75120, Uppsala, Sweden}
\affiliation{CSIR-Network of Institutes for Solar Energy, New Delhi, India}

\begin{abstract}
\noindent
We report magnetic, dielectric and magnetodielectric responses of pure monoclinic bulk phase
of partially-disordered La$_2$NiMnO$_6$, exhibiting a spectrum of unusual properties and establish that this compound is an intrinsically multiglass system with a large magnetodielectric coupling (8-20$\%$) over a wide range of
temperatures (150 - 300 K). Specifically, our results establish a unique way to obtain colossal magnetodielectricity,
independent of any striction effects, by engineering the asymmetric hopping contribution to the dielectric constant
via the tuning of the relative spin orientations between neighboring magnetic ions in a transition metal oxide system.
We discuss the role of antisite (Ni-Mn) disorder in emergence of these unusual properties.
\end{abstract}

\maketitle


Magnetodielectric compounds, whose dielectric properties depend on the applied magnetic field, and which exhibit such effects near the room temperature, hold great promise for future device applications \cite{Mostovoy2010,Kimura2010,Fiebig}.  However, such materials are rare, since the electronic origins of spontaneous magnetic and electric dipolar orderings are generally mutually exclusive \cite{NSpaldin2000}. Significant magnetodielectric effects have recently been observed in spin-spiral systems, such as TbMnO$_3$ \cite{Tokura2003} and CuO \cite{Kimura2008} at temperatures well below the room temperature and in charge-ordered systems, such as LuFe$_2$O$_4$ \cite{Subramanian2006}.  In this letter, we show that partially-disordered La$_2$NiMnO$_6$ has an array of highly interesting properties, such as a disordered ferromagnetism at higher temperatures and a reentrant spin-glass transition at lower temperatures, a relaxor-type dielectric behavior, defining this system as a rare example of an intrinsic multi-glass system in contrast to Mn-doped SrTiO$_3$ \cite{Kleeman2008,DChoudhury2011}. We show that this system is a good insulator with a colossal magnetodielectric coupling (up to $\sim$ 20$\%$) over a wide temperature range including the room temperature, and no significant magnetoresistance. We discuss the essential role played by the disorder in the form of antisite defects between Ni and Mn ions in interrelating these diverse properties. Specifically, we show that there is a substantial contribution to the dielectric constant from the mechanism of asymmetric hopping. Therefore, magnetodielectric coupling can be engineered in such systems due to the sensitivity of asymmetric hopping on the relative spin orientations of neighboring sites. This offers a new and general way to engineer magnetodielectric systems, independent of any kind of magnetostriction or electrostriction effects.

Partially-disordered La$_2$NiMnO$_6$ is synthesized in the monoclinic ($\it{P}$2$_{1}$/n) space group by the Pechini method \cite{Goodenough2003}. This is in contrast to the sample reported in ref.\cite{Subramanian2005}, which is almost fully ordered, arising from differences in details of the synthetic procedure. Rietveld refinement confirms the pure monoclinic ($\it{P}$2$_{1}$/n) La$_2$NiMnO$_6$ (LNMO) phase formation, which has two crystallographic sites for the B-site cations, Ni and Mn. Field-cooled (FC) and zero field-cooled (ZFC) DC magnetic measurements and AC susceptibility measurements were performed using an MPMS SQUID magnetometer from Quantum Design, USA. X-ray absorption spectra were collected in total electron yield mode at the I1011 beamline at the Swedish national syncrotron source MAX-lab. Dielectric constant measurements were performed in the temperature range from 10 K to 300 K and over the frequency range from 100 Hz to 1 MHz using an Agilent impedance analyzer. Dielectric constant measurements performed with sputtered gold or silver paste as electrodes gave similar results, ruling out any electrode polarization contribution. Magnetodielectric measurements were performed by recording dielectric constants in presence of a magnetic field of 2 Tesla.

\begin{figure}[t] \scalebox{1.1}{\includegraphics{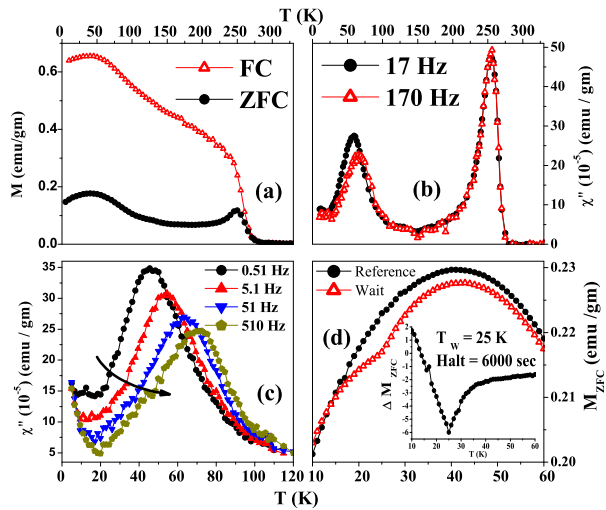}}
\caption{(Color online) (a) Field-cooled (FC) and zero-field cooled (ZFC) DC magnetization data, with an applied magnetic field (H) of 20 Oe. (b) The imaginary parts of a.c. susceptibility ($\chi''$), with H = 4 Oe. (c) same as in (b), but focusing on the low-temperature region. (d) ZFC data with and without an intermediate wait ($T_W$) at 25 K for 6000 seconds, showing distinct memory effects at $T_W$ as illustrated in the inset with a difference plot.}\label{MagRelxMono}
\end{figure}

For double perovskite structures with a general formula of $\it{A}_2$$\it{BB'}$O$_6$, it is well-known that antisite disorder with an interchange between $\it{B}$ and $\it{B'}$ sites has profound effects on physical properties, particularly magnetic properties \cite{DTopwal2006,DDSarma2000,Carlo2009,SRay2001}. Specifically in the case of La$_2$NiMnO$_6$ antisite disorder leads to Mn$^{4+}$-O$^{2-}$-Mn$^{4+}$ and Ni$^{2+}$-O$^{2-}$-Ni$^{2+}$ antiferromagnetic couplings, while the predominant Ni$^{2+}$-O$^{2-}$-Mn$^{4+}$ of the ordered structure is ferromagnetic. The saturation magnetic moment of our partially-disordered LNMO is indeed found to be 3.0 $\mu_B$/formula unit (f.u.), significantly lower than the expected 5.0 $\mu_B$/f.u. for a perfectly B-site ordered LNMO sample, indicating the presence of about 20$\%$ antisite disorder.

\begin{figure}[t] \scalebox{1.1}{\includegraphics{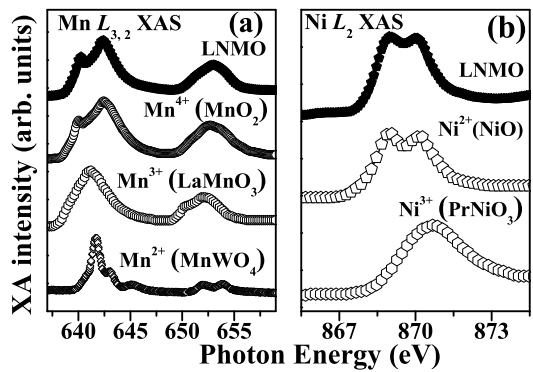}}
\caption{Mn $L_{3,2}$ (a) and Ni $L_{2}$ (b) XAS of LNMO, compared with reference spectra of compounds containing various valence states of Mn and Ni respectively, establishes that Mn exists as Mn$^{4+}$ and Ni as Ni$^{2+}$ in LNMO.}\label{XASLNMO}
\end{figure}

The magnetization ($\it{M}$) vs. temperature ($\it{T}$) curves of highly ordered LNMO shows a ferromagnetic transition at about 270 K and no indications of a subsequent magnetic transition at lower temperatures \cite{Subramanian2005}. On the other hand, the low field ZFC and FC magnetization curves (Fig.\ref{MagRelxMono}a) of our partially-disordered LNMO sample reveal a ferromagnetic transition at about 270 K and an additional anomaly at a lower temperature. Quite similar magnetization results on partially-disordered LNMO \cite{Goodenough2003} have been interpreted based on the presence of two ferromagnetic phases in the sample, namely one containing Ni$^{2+}$ - Mn$^{4+}$ ions with a T$_{c}$ $\sim$ 270 K, the other with Ni$^{3+}$ - Mn$^{3+}$ with T$_{c}$ $\sim$ 100 K. However, X-ray absorption spectroscopic (XAS) experiments, a characteristic fingerprinting tool for valence determination \cite{Debraj2010}, shown in Fig.\ref{XASLNMO} and our first-principles calculations, show the presence of only Ni$^{2+}$ - Mn$^{4+}$ ions in the LNMO sample. Thus, DC magnetization data of LNMO, combined with XAS results, clearly illustrate the inadequacy of previous hypotheses related to the origin of magnetism in LNMO, suggesting a single magnetic phase in this material. While earlier reports attributed the observation of two magnetic features to the presence of two phases with distinctly different electronic and magnetic properties, implying an inhomogeneous sample, our work, in contrast, establishes these two magnetic features as intrinsic parts of a homogeneous system in the presence of antisite defects. In order to investigate the nature of this magnetism further, AC susceptibility measurements were performed on LNMO samples. Both real ($\chi'$ not shown for brevity) and imaginary ($\chi''$) parts of AC susceptibility data, measured with applied magnetic field frequencies of 17 and 170 Hz, as shown in Fig.\ref{MagRelxMono}b, show a sharp frequency independent ferromagnetic transition around 270 K. The presence of a frequency dependent peak below 150 K, shown in detail in Fig.\ref{MagRelxMono}c for various frequencies, suggest the presence of dynamical features at low temperatures. In order to investigate whether LNMO also exhibits aging, memory and rejuvenation effects, characteristic properties of spin-glass compounds \cite{RMathieu2001,PJonsson2004}, DC memory experiments were performed. These experiments are performed by comparing the DC ZFC magnetization curves collected with and without an intermediate wait at a specific temperature ($T_{W}$). For spin-glasses, a distinct dip is observed in the difference ZFC plot, at $T_{W}$, reflecting the rearrangement of the spin configuration during the halt in the ZFC cooling, and that this equilibration, or aging, has been kept in memory \cite{Roland2010}. The dip in the difference ZFC plot at $T_{W}$ for DC memory experiments, shown in Fig.\ref{MagRelxMono}d, conclusively shows that LNMO exhibits spin-glass like dynamics at low temperatures. Hence, LNMO behaves like a re-entrant spin glass or re-entrant ferromagnet, exhibiting successive transitions from paramagnetic to ferromagnetic, and ferromagnetic to spin glass states \cite{Roland2000} with a lowering of the temperature. The existence of the low-temperature glassy state implies that the ferromagnetic state established at higher temperatures, in presence of antisite disorders, is magnetically frustrated, since the presence of antisite disorders also leads to super-exchange mediated antiferromagnetic interactions between Mn$^{4+}$-O$^{2-}$-Mn$^{4+}$ and Ni$^{2+}$-O$^{2-}$-Ni$^{2+}$. We note that, long range antiferromagnetism and short-ranged cluster glass magnetism has been discussed recently for PbFe$_{0.5}$Nb$_{0.5}$O$_{3}$. However, it was found to originate from two distinctly different magnetic phases in the sample \cite{Kleeman2010}, implying an inhomogeneous sample, similar to earlier suggestions concerning the two magnetic transitions in La$_{2}$NiMnO$_{6}$. Thus, these earlier reports are fundamentally different from the present study where we establish that a reentrant spin-glass magnetism originates within a magnetically homogenous phase.

\begin{figure}[t] \scalebox{1.1}{\includegraphics{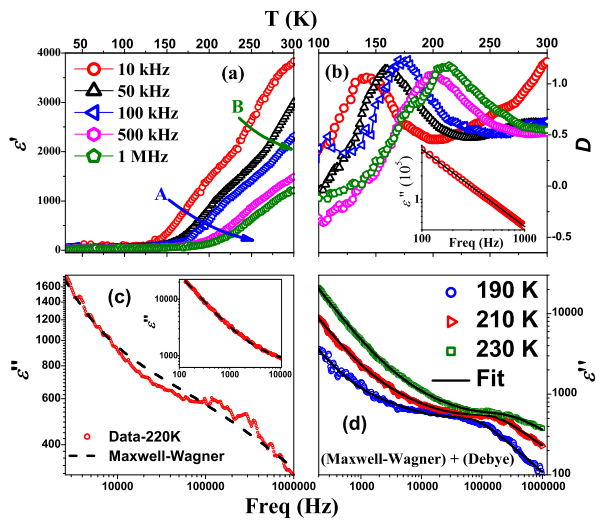}}
\caption{(Color online) (a) Dielectric constants ($\epsilon'$) for various fixed frequencies. The arrows $A$ and $B$ indicate the presence of two relaxations in the dielectric data. (b) Corresponding loss ($D=\tan\delta$) data. Inset shows the imaginary part of dielectric constant ($\epsilon''$) (symbols) at 300 K in the low frequency region; the solid line passing through the data has a slope of -1. (c)The experimental data and calculated best fit of $\epsilon''$ using only Maxwell - Wagner relaxation with the inset showing an excellent agreement in the low frequency region. (d)The experimental and best fits to the imaginary part of dielectric constant ($\epsilon''$) data of LNMO with a combination of Maxwell-Wagner relaxation and Debye relaxation.}\label{DielRelxMono}
\end{figure}

The dielectric constant of partially-disordered LNMO, measured as a function of temperature for various applied frequencies (Fig.\ref{DielRelxMono}a), show the presence of dominant frequency-dependent dielectric relaxations, though it is small and relatively frequency independent below 100 K. The presence of strong dielectric relaxations is also seen in the corresponding loss ($\tan\delta = D$) data, given in Fig.~\ref{DielRelxMono}b. To analyze the origin of observed dielectric relaxations, we first note that in the frequency range probed in these experiments, dielectric relaxations can arise from two independent mechanisms: (a) it can arise from dipolar contributions originating from asymmetric hopping of charge carriers \cite{KCKhao2004,Shu2010} (between Ni$^{2+}$ and Mn$^{4+}$ sites), in the presence of an electric field. This is expected to give rise to a Debye relaxation \cite{KCKhao2004}. (b) Dielectric relaxations can also originate from the presence of accumulated charge carriers between regions in the sample having different conductivities such as near the grain boundaries. This extrinsic source of dielectric relaxations is known as Maxwell-Wagner relaxations \cite{Hippel1966}. Maxwell-Wagner polarization mechanism can be easily identified by its characteristic $f^{-1}$ dependence of the imaginary part of dielectric constant data ($\epsilon''$) at lower frequencies \cite{Hippel1966}, whereas the Debye part of $\epsilon''$ data goes to zero. The slope of log($\epsilon''$) vs. log($f$) is found to be ($-$1) at low frequencies, as shown for 300 K in the inset to Fig. \ref{DielRelxMono}b, clearly suggesting the presence of Maxwell-Wagner relaxation. However, a Maxwell-Wagner relaxation model, fails to fit the dielectric relaxation at higher frequencies (see the comparison between experimental data and the best fit with the Maxwell-Wagner polarization in Fig. \ref{DielRelxMono}c). This suggests that additional high frequency relaxation processes exist in partially-disordered LNMO. The presence of two relaxations are also evident from Fig. \ref{DielRelxMono}a, where two dielectric relaxations, $A$ and $B$, are clearly visible, as marked. A combined relaxation mechanism, incorporating both Maxwell-Wagner and Debye relaxations, reproduces the experimental spectra well over the complete frequency and temperature ranges, as shown in Fig.~\ref{DielRelxMono}d. The Debye relaxation times extracted from the fits, correspond to an activated behavior with an activation energy of 120 meV. This is close to the activation energy of 160 meV determined from the dispersion in the peak position of loss ($D$) data for relaxation $A$ in Fig.~\ref{DielRelxMono}b. The high frequency relaxation, $A$, is thus dominated by a Debye-type relaxation while the low frequency relaxation, $B$, arises from the Maxwell-Wagner polarization mechanism.

\begin{figure}[t] \scalebox{0.9}{\includegraphics{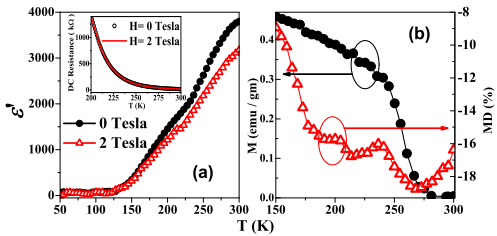}}
\caption{(Color online) (a) Dielectric constants at an applied frequency of 10 kHz in absence and presence of a magnetic field of 2 Tesla. The inset shows the absence of any DC magnetoresistance for applied magnetic fields upto 2 Tesla. (b) Temperature dependent magnetization (left axis, measured with H = 20 Oe) and magnetodielectricity (right axis).}\label{MCdataLNMO}
\end{figure}

Fig.~\ref{MCdataLNMO}a shows the dielectric constant for an external magnetic field, $\it{H}$ = 0 and 2 Tesla, establishing a strong magnetodielectric effect above approximately 100 K. Absence of any noticeable magnetoresistance effect ($<$ 0.3 $\%$), as shown in the inset to Fig.~\ref{MCdataLNMO}a, ensures that the observed magnetodielectricity is an intrinsic property of partially-disordered LNMO sample. The percentage magnetodielectricity, $MD (\%) = \frac{\epsilon_{\mathrm{2T}}-\epsilon_{0}}{\epsilon_{0}}*100$ is shown in Fig.~\ref{MCdataLNMO}b. We find similar high MD values ($\sim$16$\%$) for frequencies up to 1 MHz at 300 K. Since the Maxwell-Wagner contribution to the dielectric constant is progressively suppressed with increasing frequency, it establishes the origin of MD in partially disordered La$_{2}$NiMnO$_{6}$ to be related to the intrinsic Debye part. Interestingly, MD behavior of the present partially-disordered sample is qualitatively different from that reported earlier for the ordered sample \cite{Subramanian2005}. MD of the ordered sample vanished abruptly for $\it{T}$ $<$ 210 K, while MD of our partially-disordered sample remains substantial even at much lower temperature, e.g. being $>$ 8$\%$ at 150 K, becoming negligible only at the temperature where the system becomes magnetically glassy ($<$ 100 K) indicating an intrinsic connection between spin orientations on the two magnetic sublattices (Ni and Mn) and the magnetodielectric coupling.

We have already noted that the intrinsic part of the dielectric constant, exhibiting a Debye relaxation, arises from an asymmetric hopping between the transition metal sites in presence of the applied electric field. Moreover, this hopping is highly spin dependent, with a parallel spin orientation allowing hopping and an antiparallel orientation between the neighboring sites forbidding the possibility of hopping. We attribute the low dielectric constant and an absence of the Debye relaxation for $\it{T}$ $<$ 100 K to the random spin arrangement of the spin-glass state that suppresses any contribution from the asymmetric hopping mechanism to the dielectric constant, making the dielectric constant essentially controlled by the electronic and the phonon contributions. At higher temperatures, above the spin-glass transition, the spin-spin correlation is significantly enhanced by the application of the magnetic field, leading to an increasing MD up to about the ferromagnetic transition temperature (Fig.~\ref{MCdataLNMO}b). Thereafter the MD again decreases due to thermal disordering of the spin orientation even in presence of the magnetic field, thereby giving rise to the observed colossal magnetodielectricity, analogous to colossal magnetoresistance in the manganites. We note that unlike other models of MD, which originates from magnetostriction or electrostriction \cite{Scott2009}, our results show an unique way of engineering colossal magnetodielectricity through relative-spin orientation dependent asymmetric hopping of charge carriers between transition metal ion sites.

Using first-principles calculations based on density functional theory with a generalized gradient approximation as implemented in Quantum ESPRESSO package \cite{Mostovoy2010}, we determine structure and vibrational frequencies for the ferromagnetic (FM) and ferrimagnetic (FiM) ordering in absence and presence of antisite disorder between Mn and Ni ions in the monoclinic structure of LNMO. For configurations with different magnetic ordering and disorder, we have used a 20-atoms unit cell of the monoclinic structure. Antisite disorder is introduced by an interchange of Ni and Mn atoms in the ab-plane. Magnetic moments at all B-cations are in the same direction in the FM configuration, while those at nearest neighbours in the ab-plane are anti-parallel in the FiM configuration. This is the smallest supercell that allows us to determine effects of disorder and magnetic ordering on vibrational properties of LNMO in the monoclinic structure. We did not consider any corrections for on-site correlations or non-colinearity of spins, as our earlier works \cite{Sarma2008,UVWaghmare2009} were quite successful in capturing the physics, particularly magnetic ordering and insulating properties, of LNMO in rhombohedral structure. We find (a) the difference in energies of FM and FiM ordering reduces from 0.10 to 0.05 eV/unit with antisite disorder and (b) energy of the FM configuration with disorder is 0.08 (0.02) eV/unit higher (lower) than the FM (FiM) state of the ordered phase. Thus, the energy-scales of magnetic and chemical ordering are similar, which result in structural and magnetic frustration, and hence the observed bi-glassy behavior. Spin-phonon coupling \cite{Sarma2008} alone cannot explain the significant magnetodielectric effect, which is observed in the temperature range where phonon contribution to dielectric response is rather small. Interestingly the dielectric constant calculated from the phonon and electronic contributions is $\sim$ 46, in good agreement with the dielectric constant ($\sim$ 50) of La$_2$NiMnO$_6$ at low temperatures, where the Debye contribution becomes negligible, as already discussed. This provides a strong justification and the relevance of the present theoretical approach. Based on the details of the electronic structure, we believe that the sharp increase in dielectric constant above $\it{T}$=100 K is associated with the onset of a mechanism that involves previously mentioned asymmetric hopping in presence of the electric field amounting to a small charge transferred state (Ni$^{(2+\delta)+}$,Mn$^{(4-\delta)+}$) starting from (Ni$^{2+}$,Mn$^{4+}$) state via bridging oxygen \cite{Shu2010,UVWaghmare2010}. Since the displacement of tiny charge, is through the whole unit cell, it gives rise to a large dipole moment. We may consider perfectly ordered LNMO at a low temperature and with fully developed magnetic moment as a system with a clean gap. However, any antisite disorder will introduce mid-gap states, associated with the local deviation from the global ferromagnetic order, depending on the specific local arrangements of Ni and Mn including antisite defects. This evidently reduces the effective energy barrier (estimated to be 120 meV from experiments here), thereby making hopping more facile and increasing the dielectric constant. We should note here that the dielectric constant of the presented partially disordered sample is nearly an order of magnitude higher than the reported value of the fully ordered sample \cite{Subramanian2005}. Application of an external magnetic field, however, favors the ferromagnetic arrangement of the magnetic ions, thereby reducing the number of mid-gap states associated with local magnetic disorder, depending on the strength of the applied field. This on an average increases the energy barrier for the hopping responsible for the large dielectric constant, giving rise to the colossal negative magneto-dielectric effect observed here.

In conclusion, partially-disordered La$_2$NiMnO$_6$ exhibits a re-entrant spin-glass like magnetism, illustrated by a frustrated ferromagnetic phase established near 270 K. The system also exhibits glassy dielectric properties, establishing this system as an unique intrinsic multiglass system with a very large magnetodielectric coupling over a wide range of temperature including the room temperature (MD $>$ 16$\%$ at 300 K). We discuss the controlling influence of the partial-disorder in giving rise to these properties in conjunction with ab-initio calculations.

Authors thank Department of Science and Technology, India, Council of Scientific and Industrial Research, India, the Swedish Foundation for International Cooperation in Research and Higher Education and the Swedish Research Council (VR) and the G\"{o}ran Gustafsson Foundation, Sweden for funding.

* Also at Jawaharlal Nehru Centre for Advanced Scientific Research, Bangalore, India. Electronic mail: sarma@sscu.iisc.ernet.in

\end{document}